\documentclass[prb,superscriptaddress,aps]{revtex4}
\usepackage{graphicx}
\usepackage{dcolumn}
\usepackage{bm}

\usepackage{color}

\begin{document}


\bigskip\bigskip\bigskip

\title{Quantumlike description of the nonlinear and collective effects on
relativistic electron beams in strongly magnetized plasmas}

\author{Fatema Tanjia}
\email{tanjia@na.infn.it} \affiliation{Dipartimento di Scienze
Fisiche, Universit\`{a} Federico II and INFN Sezione di Napoli,
Complesso Universitario di M.S. Angelo, via Cintia, I-80126
Napoli, Italy}

\author{Sergio De Nicola}
\email{sergio.denicola@ino.it} \affiliation{Istituto Nazionale di
Ottica - C.N.R., Pozzuoli (NA), Italy} \affiliation{Dipartimento
di Scienze Fisiche, Universit\`{a} Federico II and INFN Sezione di
Napoli, Complesso Universitario di M.S. Angelo, via Cintia,
I-80126 Napoli, Italy}

\author{Renato Fedele}
\email{renato.fedele@na.infn.it} \affiliation{Dipartimento di
Scienze Fisiche, Universit\`{a} Federico II and INFN Sezione di
Napoli, Complesso Universitario di M.S. Angelo, via Cintia,
I-80126 Napoli, Italy}

\author{P. K. Shukla}
\email{ps@tp4.rub.de} \affiliation{Center of Advanced Studies in
Physical Sciences, Ruhr-Universität Bochum, Bochum, Germany}

\author{Du\v san Jovanovi\'c}
\email{djovanov@ipb.ac.rs} \affiliation{Institute of Physics,
University of Belgrade, Belgrade, Serbia}

\centerline {\textsf{P5.021 - 38th EPS Conference on Plasma
Physics, Strasbourg, France, 26 June - 1 July, 2011}}

\bigskip\bigskip

\maketitle

In this paper, we want to study, in a self consistent way, some
nonlinear and collective transverse effects due to the interaction
of a relativistic electron/positron beam with a magnetoactive
plasma in overdense regime, i.e., $n_0 >> n_b$ ($n_0$ and $n_b$
being the unperturbed plasma and beam densities, respectively). To
this end, in the long beam limit, we consider the self-interaction
that is produced when a relativistic charged particle beam is
travelling in the plasma exciting large amplitude plasma waves,
namely plasma wake field (PWF) excitation \cite{Chen et. al.1985}.

Hereafter, we refer to our paper \cite{P5.006} appearing in this
proceedings, as well. According to this paper, we assume that the
plasma is collisionless and cold, with ions at rest forming a
uniform background of positive charge. Furthermore, a strong
constant and uniform external magnetic field is assumed to be
acting along the z-axis, $B_0 = B_0 \hat{\mathbf{e}}_z$. We also
assume that the electron/positron beam is initially travelling
along the direction of the magnetic field with a velocity
$\mathbf{v}_b = \beta c \hat{\mathbf{e}}_z$ ($\beta \simeq 1$). We
consider a fluid model, consisting of Lorentz-Maxwell system of
equations for the \textit{beam-plasma} system. From the perturbed
Lorentz-Maxwell system, we obtain an equation that governs the
evolution of the \textit{plasma wake potential} driven by the
charged particle beam density. On the other hand, by ignoring the
longitudinal beam dynamics, we write the equation that governs the
\textit{spatio temporal} evolution of the charged particle beam
given by the thermal wave model (TWM) \cite{Fedele1991,
Fedele1992} (For details, see Ref.\cite{P5.006}). These pair of
equations can be cast as a \textit{quantumlike Zakharov system of
equations} which governs the self consistent \textit{spatio
temporal} evolution of the \textit{PWF self-interaction} of the
electron/positron beam, viz., {\small\begin{eqnarray}
&&i\epsilon\frac{\partial \Psi}{\partial\xi}=
-\frac{\epsilon^2}{2}\nabla^2_\perp\Psi-\frac{i\epsilon
k_c}{2}\hat{z}\cdot(\mathbf{r}_\perp\times\nabla_\perp)\Psi
+U_w(\textbf{r}_\perp,\xi)\Psi+\frac{1}{2}Kr^2_\perp\Psi,
\label{b5}\\
&&\left(\nabla^2_\perp- \frac{\omega^2_{pe}}{\omega^2_{UH}}
\frac{\omega^2_{pe}}{c^2}\right)U_w=
\frac{\omega^2_{pe}}{\omega^2_{UH}}\frac{\omega^2_{pe}}{c^2}
\frac{\rho_b}{n_0\gamma_0}, \label{z1}\
\end{eqnarray}}
where \small{$\omega_{UH}$} is the electron upper hybrid frequency
and \small{$\Psi = \Psi(\mathbf{r}_\perp,\xi)$} is the beam wave
function (BWF), so that its squared modulus is proportional to the
beam density, i.e., \small{$\rho_b
(\mathbf{r}_\perp,\xi)=\left(N/\sigma_z\right)|\Psi(\mathbf{r}_\perp,\xi)|^2$},
where $N$ and $\sigma_z$ are the total number of particles and the
beam length, respectively,
\small{$K\equiv(\omega_c/2\gamma_0c)^2\equiv\left(qB_0/2m_0\gamma_0c^2\right)^2\equiv
\left(k_c/2\right)^2$, $\nabla^2_\perp$} is the transverse part of
the gradient operator, {\small$  U_w = \left(A_{1z}-\phi_1\right)/
m_0\gamma_0c^2 $} is the dimensionless \textit{wake potential}
with $A_{1z}= A_{1z}(\textbf{r}_\perp,\xi)$ and $\phi_1 =\phi_1
(\textbf{r}_\perp,\xi)$ the longitudinal vector potential
perturbation and electric potential perturbation, respectively.
Here, $\textbf{r}_\perp$ is the transverse position vector, $\xi =
z-\beta c t \simeq z-ct$ plays the role of time-like variable,
$m_0$ and $\gamma_0$ are the electron/positron rest mass and the
unperturbed relativistic gamma factor of the single particle of
the electron/positron beam, respectively. In  cylindrical
coordinates, \small{$r_\perp$, $\varphi$, $\xi$}, we look for a
solution of the Zakharov-like system of the form {\small$\Psi
(r_\perp,\varphi,\xi)=
\exp\left[im\left(\varphi-k_c\xi/2\right)\right]\psi_m(r_\perp,\xi)$}
with \small{$m$} integer, taking the limiting case
\small{$\left|\nabla_\perp^2\right|\ll\omega^4_{pe}/c^2\omega^2_{UH}$.}
Let us define the transverse beam size in the form of r.m.s.,
i.e., $\sigma_m^2(\xi)=2\pi \int_0^\infty
r_\perp^2|\psi_m|^2r_\perp dr_\perp$. Under the above assumptions
and definitions, from the Zakharov-like system, we easily obtain
the following 2D Gross- Pitaevskii-type equation, viz.,
{\small\begin{eqnarray} &&i\frac{\partial
\psi_m}{\partial\xi}=-\frac{1}{2r_\perp} \frac{\partial}{\partial
r_\perp}\left(r_\perp\frac{\partial \psi_m}{\partial
r_\perp}\right)-\delta_m\mid\psi_m\mid^2\psi_m
+\left(\frac{1}{2}K_br^2_\perp+\frac{m^2}{2r^2_\perp}\right)\psi_m,
\label{s1d}\
\end{eqnarray}}
where we have introduced the following dimensionless quantities:
{\small$\xi\rightarrow\xi/\beta_0$, $r_\perp\rightarrow
r_\perp/\sigma_0$, $\psi_m\rightarrow\sqrt{\pi
m!\sigma_0^2}\,\psi_m$, $K_b=K\sigma_0^4/\epsilon^2$, $\delta_m
=n_b\sigma_0^2/n_0\gamma_0\epsilon^2m!$, $\sigma_0$} and
\small{$\epsilon$} being the initial transverse beam spot size and
the transverse emittance, respectively.
\begin{figure}[htb]
\centering
\includegraphics[width=13cm]{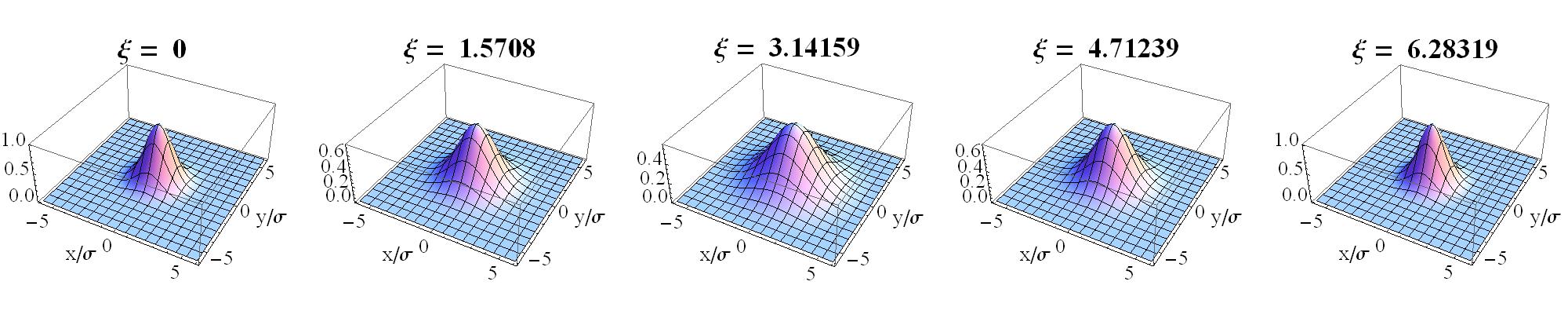}
\includegraphics[width=13cm]{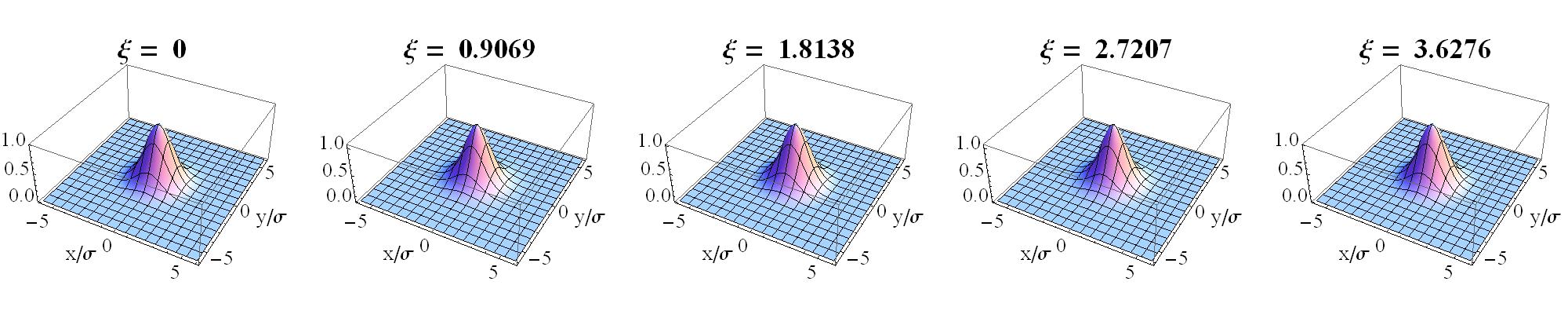}
\includegraphics[width=13cm]{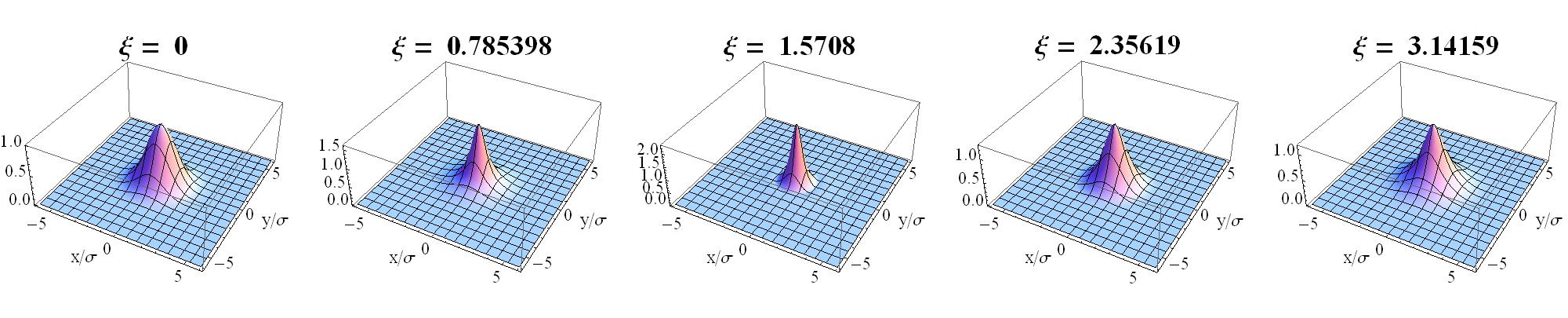}
\caption{\small{3D plots of $|\psi_m|^2$ as function of
$x/\sigma$ and $y/\sigma$ for different values of the
dimensionless time $\xi$ for $m=0$: $K_b=0.25$, $\delta_m=0.5$,
$\mathcal{A}_m=0.5$ (first row); $K_b=0.75$, $\delta_m=0.5$,
$\mathcal{A}_m=0.75$ (second row); $K_b=1.0$, $\delta_m=1.5$,
$\mathcal{A}_m=0.625$ (third row).}}\label{fig1}
\end{figure}
\begin{figure}
\centering
\includegraphics[width=13cm]{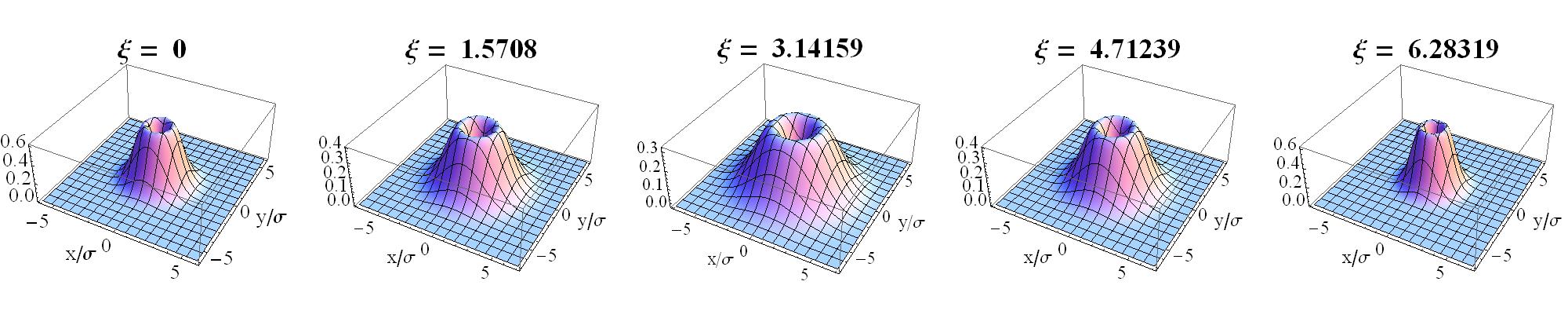}
\includegraphics[width=13cm]{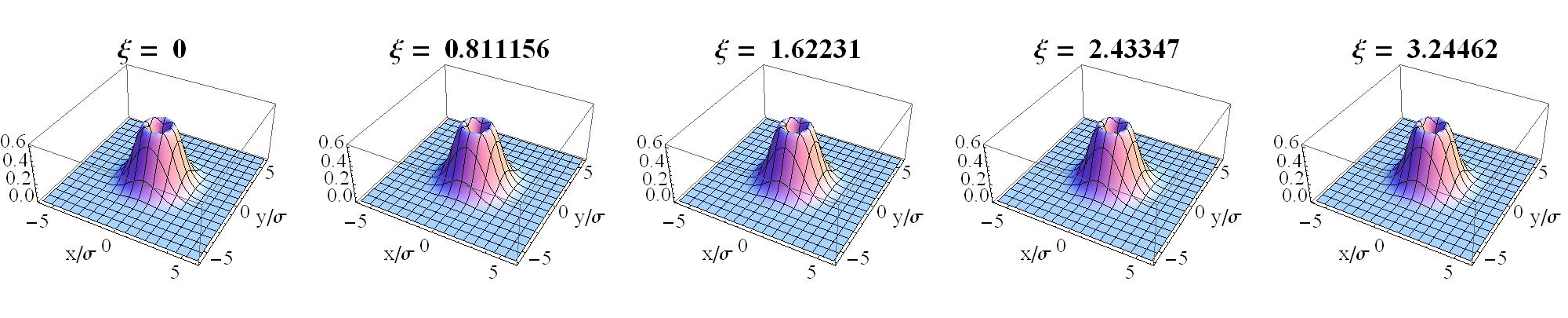}
\includegraphics[width=13cm]{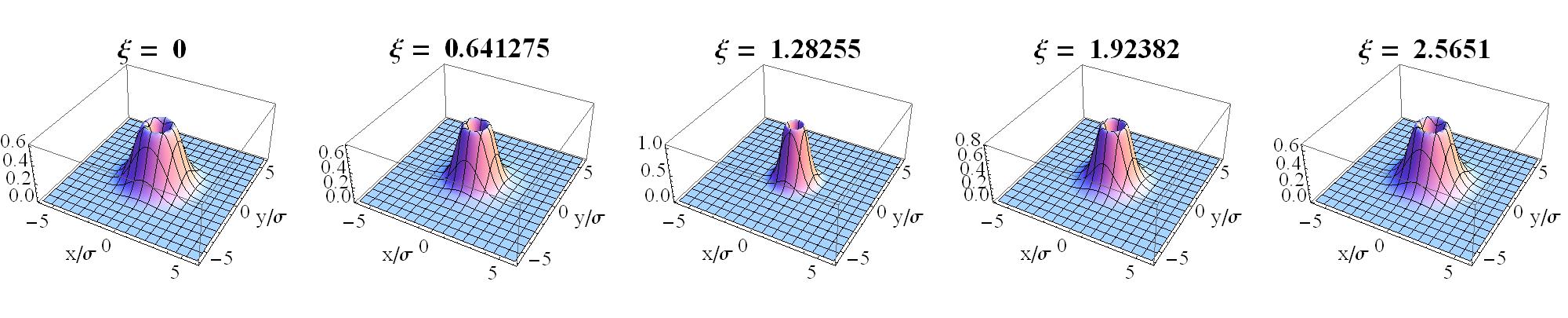}
\caption{\small{3D plots of $|\psi_m|^2$ as function of $x/\sigma$
and $y/\sigma$ for different values of the dimensionless time
$\xi$ for $m=1$: $K_b=0.25$, $\delta_m=0.5$,
$\mathcal{A}_m=1.1875$ (first row); $K_b=0.9375$, $\delta_m=0.5$,
$\mathcal{A}_m=1.1875$ (second row); $K_b=1.5$, $\delta_m=3.5$,
$\mathcal{A}_m=2.0625$ (third row).}} \label{fig2}
\end{figure}
\begin{figure}
\centering
\includegraphics[width=13cm]{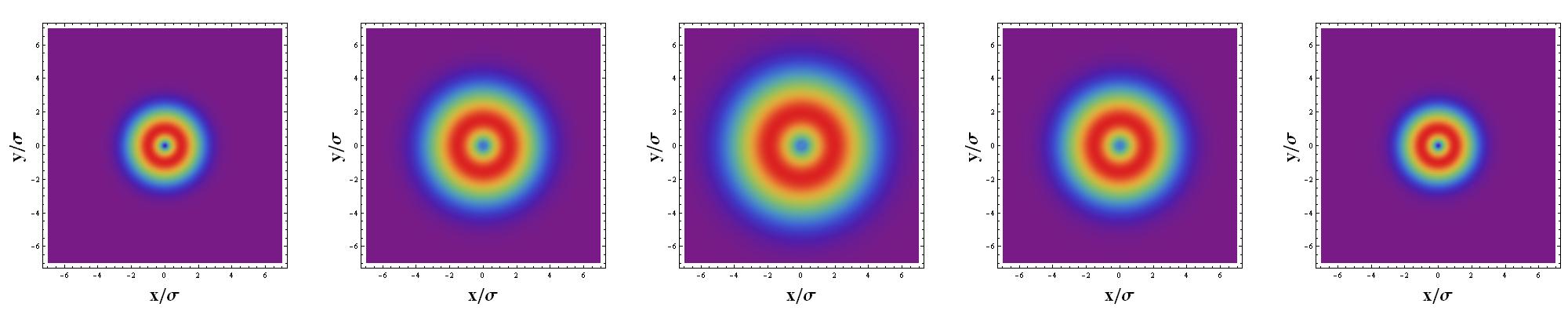}
\includegraphics[width=13cm]{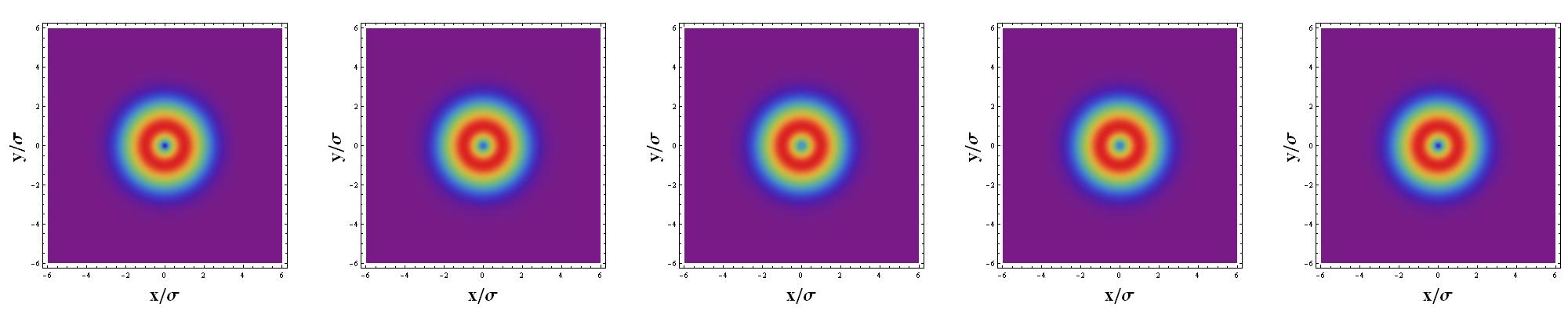}
\includegraphics[width=13cm]{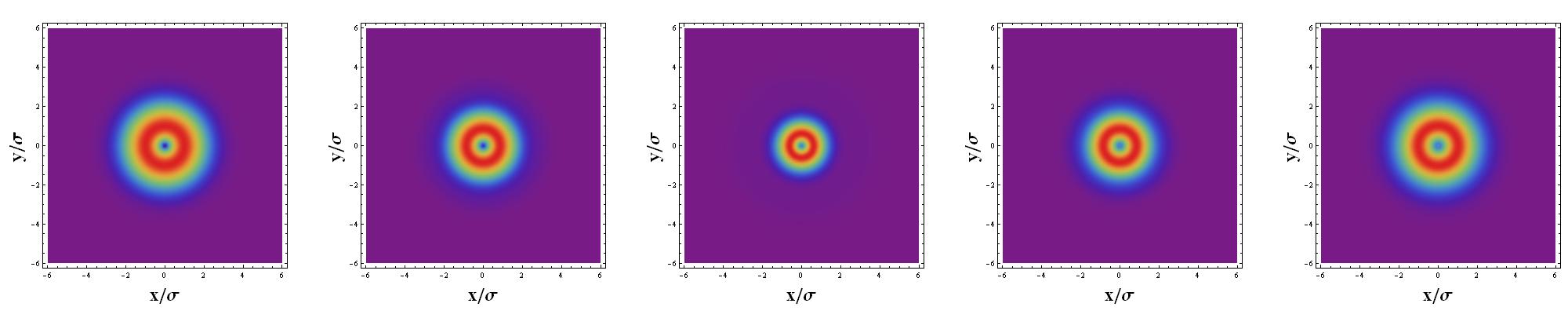}
\caption{\small{Density plots of $|\psi_m|^2$ in the $x/\sigma$,
$y/\sigma$ plane for different values of the dimensionless time
$\xi$ for $m=1$. The choice of the parameters is the one
corresponding to Figure 2: $K_b=0.25$, $\delta_m=0.5$,
$\mathcal{A}_m=1.1875$ (first row); $K_b=0.9375$, $\delta_m=0.5$,
$\mathcal{A}_m=1.1875$ (second row); $K_b=1.5$, $\delta_m=3.5$,
$\mathcal{A}_m=2.0625$ (third row).}} \label{fig2_DP}
\end{figure}
\begin{figure}
\centering
\includegraphics[width=13cm]{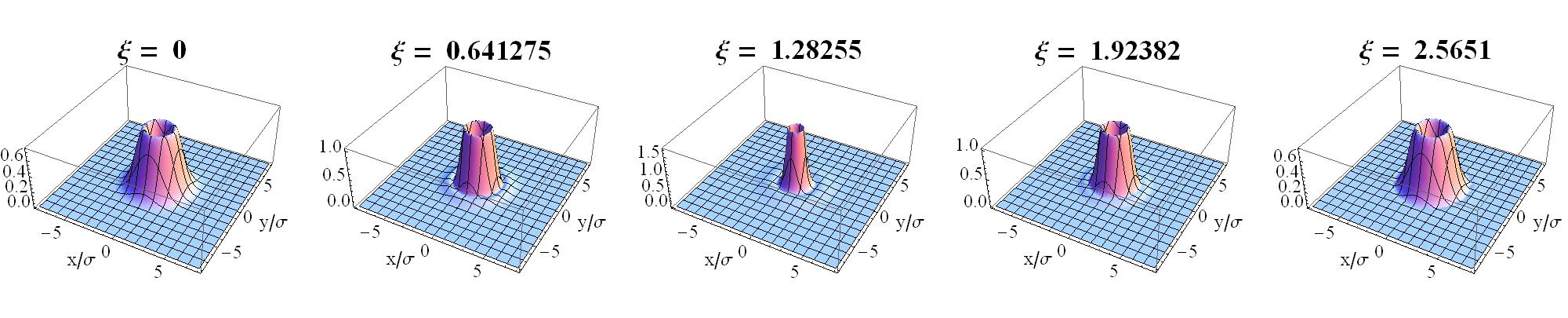}
\includegraphics[width=13cm]{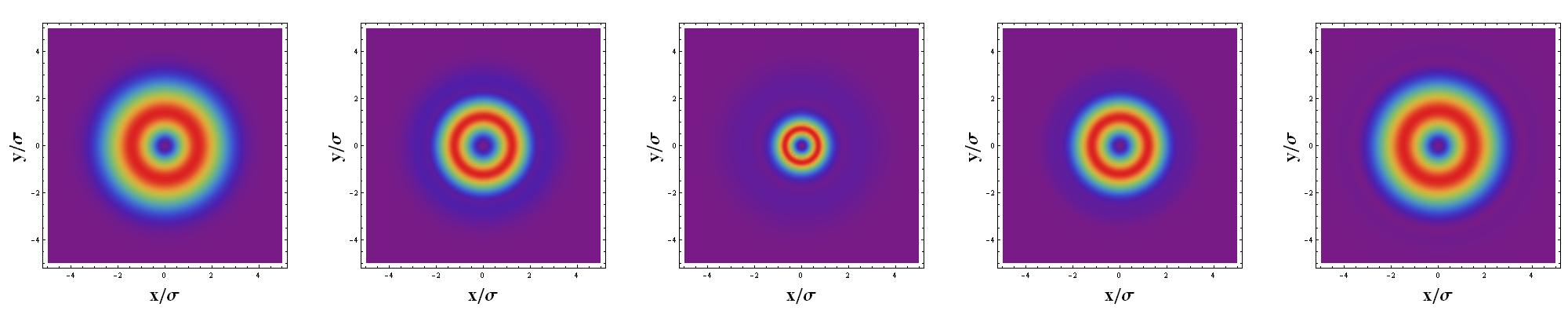}
\caption{\small{3D plots (first row) and density plots (second
row) of $|\psi_m|^2$ as function of $x/\sigma$ and $y/\sigma$ for
different values of the dimensionless time $\xi$ for $m=2$:
$K_b=1.5$, $\delta_m = 5.0$, $\mathcal{A}_m = 5.625$. }}
\label{fig3}
\end{figure}
We use the virial equation associated with eq. (\ref{s1d}) to get
the envelope equation {\small$
d^2(\sigma^{m}_{r_\perp})^2/d\xi^2+4K_b(\sigma^{m}_{r_\perp})^2
=4\mathcal{A}_m$}, where
{\small$\sigma^{m}_{r_\perp}\rightarrow\sqrt{m!}\,\sigma^{m}_{r_\perp}/\sigma_0$}
and {\small$\mathcal{A}_m=\frac{1}{2}(m+1)!(1+K_b)-\delta_m(2m)!
2^{-2(m+1)}$} is constant of motion. Note that, from the envelope
equation, the matching condition for the equilibrium transverse
beam spot size, {\small$\sigma_{eq}^m$, is $K_b(\sigma_{eq}^{m})^2
=\mathcal{A}_m$}.\\
A preleminary numerical analysis has been carried out by solving
eq. (\ref{s1d}) assuming the initial normalized BWF (density
profile) as
{\small$\psi_m(r_\perp,0)=r_\perp^m\exp\left(r_\perp^2/2\right)$}.
The spatio-temporal evolution of $|\psi_m|^2$ has been
investigated for different values of $m$, $K_b$, and $\delta_m$,
at \small{$\xi= 0$, $0.25 T$, $0.5T$, $0.75T$, $T$,} where
$T=\pi/\sqrt{K_b}$. For both $m = 0$ and $m = 1$, when the
matching condition of the envelope equation is satisfied, the
profile is practically unchanged (see the second row of Figures 1,
2 and 3, respectively). This predicts the existence of nonlinear
coherent states (sometimes called 2D solitons). Furthermore, due
to the strong nonlinearity, the effect of \textit{beam halo} has
been observed for $m = 2$, as displayed by both 3D and density
plots in Figure 4. Due to the interplay between the strong
transverse effects of the plasma wake field (collective and
nonlinear effects) and the magnetic field, envelope oscillations
with weak and strong focusing and defocusing have been observed
for $m=0$ (see the first and the third row of Figure 1), $m=1$
(see the first and the third row of Figures 2 and 3, respectively)
and $m=2$ (see first and second row of Figure 4). Finally, the
existence of vortices (effect of the orbital angular momentum due
to the external magnetic field) are clearly shown in Figures 2,3
and 4, respectively.\\
The present investigation seems to be useful for the plasma-based
focusing schemes to be employed in the final focusing stages of
linear colliders as well as for manipulating relativistic
electron/positron beams that suggests the new fields of nonlinear
and collective singular electron optics.

\end{document}